# GAMMA RAY BURSTS: RECENT RESULTS OBTAINED BY THE SWIFT MISSION

GUIDO CHINCARINI

*Università degli Studi di Milano Bicocca & Osservatorio Astronomico di Brera*
*Via E. Bianchi 46, Merate (LC), Italy*

*ON BEHALF OF THE*
Swift TEAM

In the introduction we give the main characteristics of the Swift mission outlining how the design was driven by the science goals and by the heritage we got from the Italian – Dutch satellite Beppo SAX. We show some of the new characteristics of the X – ray light curves that became evident soon after we obtained the first set of data. In addition to the early phase steep slope afterglow discovered by Swift, we discuss the frequently observed GRB flares and the first localization of a short burst.

## 1. Introduction

With the launch of the Swift satellite, Gehrels et al.[1] a new era started not only for the study of the Gamma Ray Bursts but also for various fields of Astrophysics covering stellar evolution, relativistic stars and merging, relativistic jets and cosmology. Indeed the brightness of the GRB events supports the hope to detect these objects at very high redshifts and near the re-ionization epoch soon after the Universe gets out of the Dark Age. The design of the mission was carried out with some of these goals in mind and to some extent tailored about the discoveries made by Beppo- SAX, Costa et al.[2], Van Paradijs et al.[3], Frail et al.[4].

The main ingredients of the mission design were multi-wavelength coverage, state of the art instrumentation with good resolution and sensitivity, high speed re-pointing capability of the spacecraft and a fast and reliable communication system. These tasks were satisfied by a payload composed by three instruments: The Burst Alert Telescope (BAT), the X ray Telescope (XRT) and the Ultraviolet Optical Telescope (UVOT), by the Malind ASI ground station and TDRSS.

### 1.1. *Burst Alert Telescope (BAT)*

BAT, sensitive in the energy range 15 – 150 keV for imaging and with non-coded response reaching 350 keV, monitors the sky with its large field of view of about 1.4 steradians (100° x 60°). Within few tens of seconds of detecting a burst the on board computer will calculate the initial position with an accuracy of better than 4 minutes of arc and decide, after accounting also for all the pointing constraints, whether the burst merits a spacecraft slew. Assuming the burst is worthy, it also sends the position to the spacecraft for pointing the object and setting it on axis of XRT and UVOT. In various cases we were able to begin observations with the Narrow Field Instruments (NFI) within less than 60 seconds since the BAT alert.

### 1.2. *X ray Telescope (XRT)*

XRT is sensitive in the energy band 0.2 – 10 keV, has an effective area of 110 $cm^2$ and has the capability to estimate an accurate position, accuracy better than 5", in less than 100 seconds. Details of this instrument can be found in Burrows et al[5]. An important characteristic of XRT is the capability to switch automatically between different modes of operations in order to have the capability to deliver accurate photometry of the source over a range of about seven order of magnitude (from $2\ 10^{-14}$ erg $cm^{-2}$ $s^{-1}$ to about $8\ 10^{-7}$ erg $cm^{-2}$ $s^{-1}$ ) . This is a requirement for observing bright and rapidly fading sources. Depending on the flux (count rate) of the source, therefore, the instrument switch from Image mode (IM) that is used at the very beginning (w/o spectral information) to determine the source position, to either Photo Diode mode (PD), used for very bright sources with a time resolution of 0.14 ms and no imaging information or Window Timing mode (WT) with a 1.8 ms resolution and space resolution in only one axis. For faint objects the instrument uses the Photon Counting mode (PC) with a time resolution of 2.5 seconds. This is the standard XRT mode. The quality of the X ray telescope is really excellent and on fields in which we did follow a burst for a long time we discovered we can reach a sensitivity limit of about $3\ 10^{-16}$ erg $cm^{-2}$ $s^{-1}$ (0.5 – 2 keV). These characteristics coupled to a good space resolution (18 arcsec HEW at 1.5 keV) over a 23 arcmin gave us the possibility of new discoveries in the field.

.

It is also known, however, that we lost the Thermo-Electric Cooling (TEC). Consequently the temperature of the CCD rather than being kept at -100 °C as planned, is oscillating between -85 and -45 °C since it uses only the passive cooling. Tests and calibrations that were carried out after that lost demonstrated we did not loose any sensitivity or resolution. On the other hand we must compensate with a more careful planning of the observations, to keep the CCD as cold as possible, and with more sophisticated calibrations and analysis.

**1.3.** *Ultraviolet Optical Telescope (UVOT)*

UVOT, Mason et al.[6], is a 30 cm aperture Ritchey – Chretien telescope with a 17x 17 arcmin FOV. It uses a CCD with a set of filters and grisms working in the wavelength range 1700 – 6500 Å with a space resolution of 0.5"x 0.5" on the sky. It can reach a position accuracy of 0.3 arcsec and a limiting magnitude (5 σ accuracy) B= 24.0 in 1000 s. It is essential in order to have a statistical significant sample not affected by the weather conditions and different size telescopes.

**1.4.** *The Transmission network*

Fast communication is essential for monitoring the observations of rapidly varying events and for alerting the ground based telescopes or other space facilities as soon as possible. To this end we use the TDRSS for immediate down and up loading of data and commands and the Italian ASI ground segment at Malindi for all ordinary operations.

**2.** *Observations – First light*

BAT performed quite well since the very beginning delivering excellent light curves in the 15 – 150 energy band and up to 350 keV. The first image of XRT, Cas A, is reproduced in Figure 1 next to an image obtained with the XMM telescope for comparison. The space resolution is excellent.

**3.** *Early science results*

This Cividale meeting came at a very interesting time because we had the possibility to show the discovery of the main characteristics of the light curves observed by XRT and the accurate localization of the first short burst, GRB050509. These results, at the time of writing, have been confirmed with the detailed analysis of many XRT light curve and with the discovery of the short burst GRB050709 (HETE) and GRB050724. In addition Swift discovered the high z GRB050904 (z = 6.29) long after the meeting and this we will not discuss in this proceedings.

### 3.1. *XRT light curves*

The analysis of the following light curves, GRB011121, GRB050126, GRB050219A, GRB050315 and GRB050319, showed since the very beginning the main characteristics that have been described in a more organic way by Chincarini[6] et al. and by Nousek[7] et al.

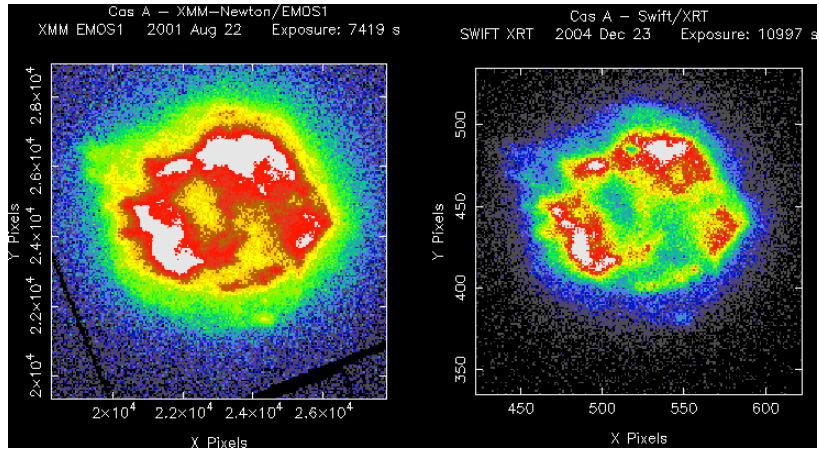

Figure 1. Left: Cas A obtained by XMM on August 2001 with an integration of 7419 seconds. Right Cas A observed by XRT on December 2004, during the activation phase, with an exposure of 10997 seconds.

It is clear from Figure 2 that the basic morphology is given, soon after the prompt emission, by a fast decay of the afterglow, see also Tagliaferri[8] et al., followed by a milder decay steeping down after a while. This will define the type 1 light curve. In one case, GRB050319 in Figure 2, the initial slope is much higher (about -6) but that is almost certainly due to an erroneous BAT trigger. Counting the time ($t_0$) from the beginning of the second BAT burst (in this GRB the prompt emission is composed by two bursts separated by about 137 seconds) the light curve of GRB050319 is exactly superimposed to the light curve of GRB050315. A type 2 light curve seems also to occur. This starts off with a mild slope soon after the prompt emission to steep down later on, see also Figure 3 in Chincarini et al.. The type 1 light curve is shown in Figure 2 by the dashed dotted line that represents the mean of the various type 1 light curves seen in that plot. GRB 050401 is on the contrary the prototype of a type 2 light curve.

The X-ray spectrum remains unchanged when measured before and after the break, GRB050319 being an exception, and it is very similar for the various burst with an energy index $\beta \sim 1.1$. The energy measured in the afterglow changes considerably for the various bursts and is in between 1.6% to 40% of the energy emitted by the prompt emission.

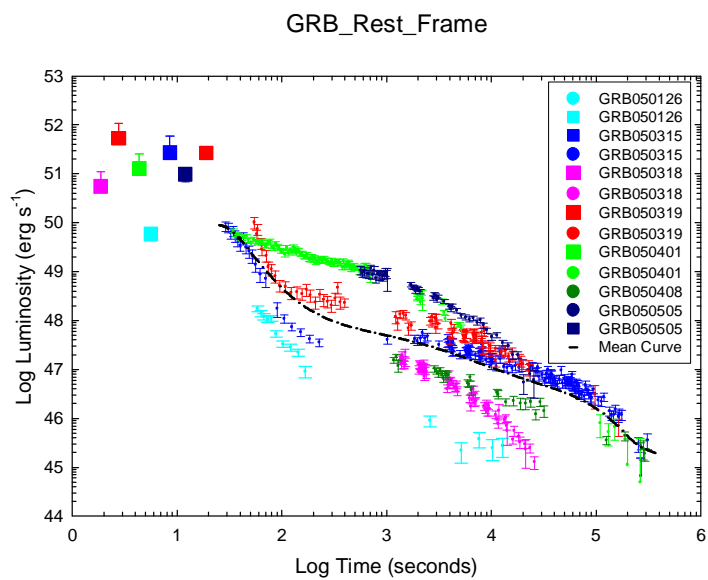

Figure 2. The first set of XRT light curves for which the redshift has been measured. There is a clear indication of continuity between the BAT and the XRT emission, for details see Chincarini et al.

The observations of GRB011121, GRB050219A, 050406 and 050502B showed very clearly that superimposed to these basic features we may observe flares, some of which are very energetic. A beautiful example of type 1 with flares is shown in Figure 3 (note however that this was obviously not shown at the time of the meeting) in which we wrote for convenience also the measured fluence.

Light curves with large flares have afterglow fluence (XRT) comparable to that of the prompt emission (BAT).

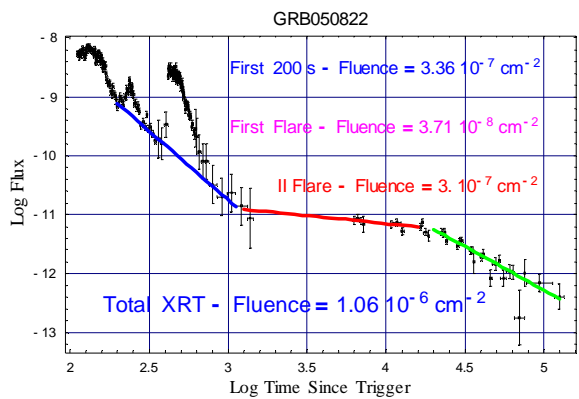

Figure 3. The XRT light curve illustrate properly all the characteristics described in the text. The fluence measured by BAT in the band 15 – 350 keV is $3.1 \times 10^{-6}$ cm$^{-2}$.

While the change of slope at the first break may need further modeling to be fully understood, the second break of the type 1 light curve (or the first of the type 2 light curve) occurs when the relativistic beaming angle reaches, because of the decreasing of Lorentz factor of the relativistic shells, the value of the jet angle, Figure 4.

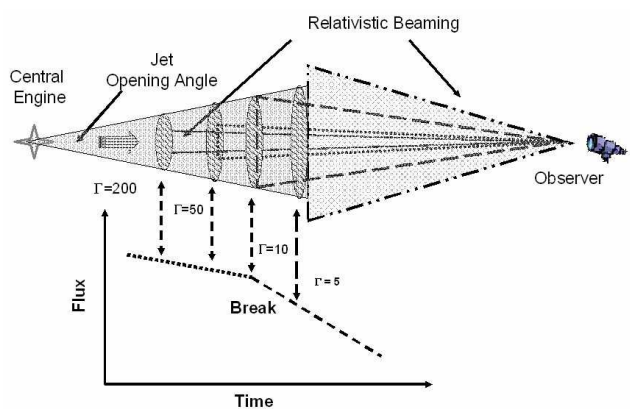

Figure 4. In this sketch the shell slow down from $\Gamma=200$ to $\Gamma=8$ and equals the Jet opening angle when $\Gamma = 10$ ($\theta = \Gamma^{-1}$). At that point the light curve shows a break and is steeping (the value of the Lorentz factors are only for illustration and are not real).

## 4. *The localization of short gamma-ray bursts*

The duration of gamma ray bursts is known to have a bimodal[9], short and long, distribution and BATSE showed that short bursts were harder[10]. Short bursts, duration less than 3 s, have been elusive for decades because of the difficulty in getting an accurate location soon after the alert. On May 5th BAT detected a burst, Gehrels[11] et al. (see error circle top left of Figure 5), with a fluence of 9.5 ± 2.5 $10^{-9}$ erg cm$^{-2}$ in the band 15 to 150 keV, right panel of Figure 5. Automatically the spacecraft re-pointed on the BAT coordinates and XRT was collecting data 62 s after the burst trigger.

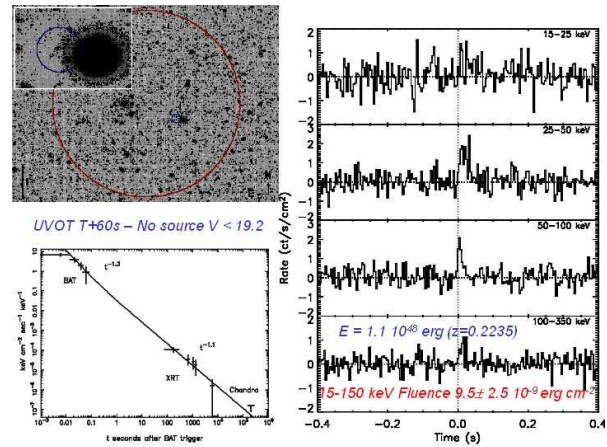

Figure 5.The first localization of a short gamma-ray burst. Top left: The field of the sky with the error circle of BAT and in the inset the E1 galaxy with the error circle of XRT. Left bottom: the XRT light curve including a late observation by Chandra. Right: the BAT light curve at different energies.

XRT collected 11 photons and gave the position with an error circle of only 9.8" (inset top left frame of Figure 5). The X-ray light curve (0.3 – 10 keV) (we can now show the complete light curve, bottom left panel of Figure 5, including the Chandra data) shows a classical decay power law with α = -1.1. We, and other groups, immediately imaged the field with medium large telescope (TNG,VLT, SUBARU, Keck etc.) and we noticed that the XRT error circle included the second brightest galaxy (E1) of the cluster of galaxies NSC J123610+295901. The cluster has a redshift z = 0.225. Within the XRT error circle we observed also various faint galaxies, likely at high redshift, and a somewhat brighter spiral.

We opted for the bright E1 elliptical as the host galaxy because of the extremely low probability that a random burst would have to be located next to a bright nearby galaxy. We were also guided in a small part by some of the possible theoretical predictions[12, 13, 14] (without having a prejudice however).  Indeed we discussed merging of relativistic stars (NS-NS, BH–NS), SGR, peculiar collapsars etc. But the idea of NS – NS or NS – BH was very appealing and in agreement with the age of the parent galaxy and the outskirt location due to the rather large peculiar velocity gained by the binary system as a consequence of the kick off received after the supernova event. This interpretation agrees well also with the estimated energy release assuming the burst is at the distance of the galaxy, $E_{iso}$=1.1 $10^{48}$ erg, and with the age needed for such a system to form and merge. Following this event that has been reported at the meeting in a preliminary form, the burst detected by HETE, GRB050709, and that detected by Swift, GRB050724, whose detailed light curve was followed by XRT, fully confirmed the above interpretation and we have now, and possibly more in the coming months, the data needed to constraint and understand the progenitor.

**Acknowledgments**

The work is supported in Italy by funding from ASI on contract number I/R/039/04, at Penn State by NASA contract NAS5-00136 and at the University of Leicester by PPARC contract number PPA/G/S/00524 and PPA/Z/S/2003/00507. We acknowledge in particular all those member of the Swift Team at large who made this mission possible. This goes from the building of the hardware, the writing of the software, the operation at the Mission Operation Centre and the performance of the ASI ground segment at Malindi, Kenya.